\documentclass[journal,comsoc]{IEEEtran}

\makeatletter

\def\ps@IEEEtitlepagestyle{%
  \def\@oddfoot{\mycopyrightnotice}%
  \def\@evenfoot{}%
}

\def\mycopyrightnotice{%
  {\footnotesize 979-8-3503-9591-4/24/\$31.00~\copyright~2024 IEEE\hfill}
  \gdef\mycopyrightnotice{}
}

\IEEEoverridecommandlockouts

\usepackage{cite}
\usepackage{amsmath,amssymb,amsfonts}
\usepackage{algorithmic}
\usepackage{graphicx}
\usepackage{textcomp}
\usepackage{url}
\def\BibTeX{{\rm B\kern-.05em{\sc i\kern-.025em b}\kern-.08em
    T\kern-.1667em\lower.7ex\hbox{E}\kern-.125emX}}

\begin{document}

\title{Enabling Sustainable Urban Mobility: The Role of 5G Communication in the Mobilities for EU Project\\}

\author{
	\IEEEauthorblockN{
		Shangqing Wang\IEEEauthorrefmark{1},
		Christopher Lehmann\IEEEauthorrefmark{1},
            Rico Radeke\IEEEauthorrefmark{1},
		and
		Frank H. P. Fitzek\IEEEauthorrefmark{1}\IEEEauthorrefmark{2}\\
	}
	\IEEEauthorblockA{
		\IEEEauthorrefmark{1} Deutsche Telekom Chair of Communication Networks, TU Dresden, Germany\\
		}
	\IEEEauthorblockA{
		\IEEEauthorrefmark{2} Centre for Tactile Internet with Human-in-the-Loop (CeTI)\\
		}
	E-mails: \{first name.last name\}@tu-dresden.de
}

\maketitle

\bstctlcite{bibliography:BSTcontrol} 

\begin{abstract}
This paper examines the role of 5G communication in the Mobilities for EU project, a collaborative initiative involving 29 partners and 11 pilots aimed at revolutionizing urban mobility through electrification, automation, and connectivity. Focusing on Dresden as a Lead City, we explore the integration of 27 innovative solutions, including autonomous freight transport, eBuses, and charging robots, using a 5G communication network as the central framework. We analyze how 5G enables seamless connectivity and real-time data processing across diverse technologies, fostering interdependencies and synergies. This approach not only provides a cohesive understanding of the project's scope but also demonstrates 5G's critical role in smart city infrastructure. We evaluate the anticipated impact on sustainability metrics such as air quality, noise levels, CO2 emissions, and traffic congestion. The paper concludes by discussing challenges and strategies in leveraging 5G for comprehensive urban mobility solutions and its potential impact on future smart city developments.

\end{abstract}


\section{Introduction}
Smart cities represent the forefront of urban innovation, leveraging advanced technologies to enhance sustainability, efficiency, and quality of life~\cite{sharma2023smart}. The European Green Deal sets ambitious targets to reduce greenhouse gas emissions from transport by 90\% by 2050, emphasizing sustainable, smart, and resilient mobility~\cite{EUTransportReport2023}. A key component in achieving these goals is the implementation of 5G communication technology, which enables various smart city applications, including intelligent transportation systems (ITS), energy management, and public safety solutions~\cite{guevara2020role}. 

5G wireless communication systems provide massive system capacity with high data rates, very low-latency, and ultra-high reliability, making them particularly suitable for smart city applications~\cite{shehab20215g}. The 5G network architecture, comprising enhanced Mobile Broadband (eMBB), massive Machine Type Communications (mMTC), and ultra-Reliable Low Latency Communications (uRLLC)~\cite{series2015imt}, provides a comprehensive framework for addressing various urban mobility challenges. 

This paper is an early stage concept paper, which explores how 5G technology enables and enhances the diverse pilots within the Mobilities for EU project, particularly concerning sustainability indicators. Launched in January 2024, the Mobilities for EU project aims to revolutionize urban mobility through innovative technologies, demonstrating the feasibility of new mobility concepts for passenger and freight transport. The initiative involves 29 partners, 11 pilots, and 27 innovative solutions, focusing on electrification, automation, and connectivity, with Dresden playing a crucial role as a Lead City implementing 14 solutions~\cite{mobilitiesforEU}.

We utilize the CIVITAS process and its impact evaluation framework to systematically assess the interventions introduced by the Mobilities for EU project. This paper examines how 5G technology addresses the challenges of integrating diverse mobility solutions and contributes to achieving the climate goals set by the European Green Deal.

\textbf{Structure of the Paper}: The paper is organized as follows: \textbf{Background} outlines the European Green Deal, smart city concepts, and urban mobility challenges. \textbf{Proposed Smart City Solutions} examines the core aspects of 5G and its applications in urban mobility. \textbf{5G Communication Network as an Enabler} analyzes the project with a focus on implementations in Dresden's Ostra district, detailing how various solutions integrate through 5G networks. \textbf{Sustainability Impact and Evaluation} assesses the impact on sustainability metrics through a case study on Vehicle-to-Grid (V2G) technology, highlighting its role in balancing energy supply and demand and the broader implications of 5G-enabled urban mobility solutions.

We conclude with an outlook on how these technologies can shape the future of smart city developments, highlighting both the opportunities and challenges in leveraging 5G for comprehensive urban mobility solutions. 

\section{Objectives and Scope}

Current research on 5G in smart cities focuses primarily on environmental sustainability (42\%), followed by economic (37\%) and social (21\%) dimensions. Within these categories, energy efficiency (20\%), power consumption (17\%), and cost (15\%) are the most studied aspect~\cite{shehab20215g}. This paper addresses these dimensions by focusing on the role of 5G in the Mobilities for EU project. The specific objectives are: \textbf{Connect and Enhance Pilot Projects:} Demonstrate the importance of 5G technology in linking various pilot projects under the Mobilities for EU initiative. 5G's high-speed, low-latency connectivity is crucial for real-time data processing and reliable communication across mobility solutions. \textbf{Overview of Dresden Pilots:} Provide a detailed overview of the pilot projects in Dresden, highlighting how these initiatives are integrated through 5G technology, including applications like autonomous freight transport, eBuses, and charging robots. \textbf{Scope of the Study within Dresden's Ostra District:} Focus on the implementation and impact of 5G-enabled solutions within the Ostra district, exploring how 5G facilitates the integration of smart city technologies to address urban mobility challenges.

The Scope of this study includes: \textbf{Technological Integration:} Examining how 5G serves as a unifying technology for diverse smart city solutions. \textbf{Sustainability Impact}: Evaluating the environmental benefits achieved through 5G-enabled smart mobility solutions. \textbf{Social and Economic Dimensions:} Analyzing how 5G contributes to social inclusivity and economic viability in smart city projects. \textbf{Real-World Applications:} Presenting case studies from the Mobilities for EU project that highlight successful implementations and areas for improvement.

By addressing these objectives, the paper aims to provide a comprehensive understanding of the critical role of 5G communication technology in advancing smart urban mobility and achieving sustainable development goals.

\section{Background}
The European Union recognizes cities as vital in the transition to a climate-neutral Europe, with the \textbf{European Green Deal} aiming for carbon neutrality by 2050~\cite{boeri2021climate}. Key urban mobility objectives include reducing transport greenhouse gas emissions by 90\% by 2050, deploying 30 million zero-emission vehicles by 2030, and implementing sustainable urban mobility plans~\cite{EUGreenDeal, EUTransportReport2023}. Achieving these goals requires innovative urban mobility solutions.

\textbf{Smart city concepts} that leverage advanced technologies like 5G are critical for these targets~\cite{TUDresdenSmartCity}. 5G enhances connectivity, low latency, and capacity, enabling real-time traffic management, integration of zero-emission vehicles into smart grids, and efficient public transportation systems~\cite{guevara2020role, shehab20215g}.

Aligned with these innovations, the Mobilities for EU project involves 29 partners and 11 pilots to transform urban mobility through electrification, automation, and connectivity. This initiative aims to create sustainable and livable cities~\cite{mobilitiesforEU}.

Dresden serves as a Lead City in this project, implementing 14 innovative solutions that exemplify its objectives. A key element is the implementation of 5G technology, which supports various smart city applications such as ITS and energy management. 5G enables seamless connectivity for real-time data processing among diverse mobility solutions, fostering interdependencies and enhancing urban mobility effectiveness~\cite{mobilitiesforEU}.

\section{Proposed Smart City Solutions}
The Ostra district in Dresden features 28 sports arenas and associated parking facilities, making it a significant hub for events and a vital contributor to the local economy and community well-being. The Mobilities for EU project is developing innovative concepts for smart management in mobility, logistics, and energy, with a strong emphasis on sustainability. Key pilot projects include:

\paragraph*{Mobile Charging Robot for EVs}
Drivers in the Ostra district can request mobile charging robots via an app to charge their electric vehicles (EVs) based on specific needs, such as current battery level and available charging time. These robots efficiently move between vehicles, maximizing the utilization of charging stations compared to fixed terminals.

\paragraph*{Automated Connected Driving}
Intelligent sensors at road junctions detect vulnerable road users (VRUs) and visitor flows, transmitting this information via V2X communication to enhance safety. Autonomous shuttles provide efficient transport and parking solutions, utilizing teleoperated driving to manage people and freight.

\paragraph*{Electrification of Public Transport and Bidirectional Charging} 
Electric public transport vehicles are deployed and evaluated for performance. Additionally, electric vehicles with bidirectional charging capabilities are implemented as mobile energy storage units, allowing excess green energy to be stored and fed back into the grid for stabilization.

\paragraph*{Connected Platform for Services and Information} 
An AI-supported data platform monitors the Ostra district's utilization, providing insights into traffic flows, parking, and charging needs. By integrating existing databases, it offers intermodal transportation options and comprehensive billing models including bookings for event tickets and hotels.

\paragraph*{Expansion of the required Infrastructure} 
A private 5G network will be established to support mobile robots and teleoperated driving, providing reliable communication and optimized processing. Power grid enhancements facilitate efficient green energy transfer to electric vehicles through bidirectional charging stations and terminals.

\section{5G Communication Network as an Enabler}


\subsection{Key Enabling Technologies for 5G in Smart Urban Mobility}
\label{sec:KeyEnabler}
To fully realize the potential of these 5G components in urban mobility, several enabling technologies such as private networks and edge computing are critical. These technologies collectively enhance the capabilities of 5G networks to support smart urban mobility solutions~\cite{Accenture2021}. 

\paragraph*{Private Networks}
Private 5G networks operate in a licensed frequency domain, separate from public frequencies. Operators typically use dedicated infrastructure, allowing network performance to be tailored to specific use cases while ensuring high levels of data privacy, integrity, and security. These geographically limited networks reduce attack risks, providing greater reliability and security for critical urban mobility operations, such as autonomous vehicle control and traffic management systems.

\paragraph*{Edge Computing}
Edge computing facilitates real-time data processing for applications like traffic signal optimization and collision avoidance, essential for smart urban mobility. By leveraging local computing power within the communication network, applications can run efficiently, leading to bandwidth savings and enhanced resiliency, reliability, and latency. This approach also reduces costs and increases the flexibility of deploying 5G infrastructure across cities, enabling broader coverage for mobility solutions~\cite{Accenture2021}. Together, these technologies enhance the capabilities of 5G networks to support smart urban mobility solutions.

\vspace{-1em}
\subsection{Major Usecases in 5G} 
The 5G network architecture comprises three key usecases that can be used to categorize and analyze the various pilots in the Mobilities for EU project.
\subsubsection{Enhanced Mobile Broadband (eMBB)}
eMBB provides high data rates and increased capacity, which are essential for applications requiring large amounts of data transmission and real-time processing, potentially reducing congestion and emissions~\cite{morocho2019machine, imran2019energy, shehab20215g}. Examples of solutions under this category include:
\textbf{Real-time Video Analytics for Traffic Management:} High-definition cameras and video analytics tools that monitor traffic flow and detect incidents in real-time. This helps in reducing traffic congestion and improving road safety.
\textbf{High-definition Mapping for Autonomous Vehicles:} Detailed and up-to-date maps that support the navigation and operation of autonomous vehicles, enhancing their efficiency and safety.
\subsubsection{Massive Machine Type Communications (mMTC)}
mMTC supports a large number of connected devices, making it ideal for IoT applications in smart cities, facilitating smart parking systems and environmental monitoring, which can improve air quality and energy efficiency~\cite{shehab20215g}. Examples of solutions under this category include:
\textbf{IoT Sensors for Environmental Monitoring: }Deployment of sensors to monitor air quality, noise levels, and other environmental parameters. This data is crucial for assessing the environmental impact of urban mobility solutions and implementing measures to improve air quality.
\textbf{Smart Parking Systems:} IoT-enabled parking sensors that provide real-time information on available parking spaces, reducing the time and fuel spent searching for parking and thus lowering emissions.
\subsubsection{Ultra-Reliable and Low Latency Communications (uRLLC)}
uRLLC ensures minimal latency and high reliability, which are critical for applications requiring real-time communication and control ~\cite{morocho2019machine, shehab20215g}. Examples of solutions under this category include:
\textbf{Vehicle-to-Everything (V2X) Communications}: Real-time communication between vehicles and infrastructure to enhance traffic safety and efficiency. This includes applications like collision avoidance systems and traffic signal optimization.
\textbf{Remote-Controlled Vehicles:} Use of 5G to remotely control vehicles for tasks such as autonomous freight transport. This reduces the need for human drivers and enhances the efficiency of logistics operations.

\vspace{-1em}
\subsection{Pilot Categorization} 

\begin{table}[h]
\centering
\caption{Mapping of Communication Requirements}
\label{table:mapping}
\resizebox{\columnwidth}{!}{%
\begin{tabular}{|l|l|l|l|}
\hline
\textbf{Smart City Solution}        & \textbf{uRLLC} & \textbf{mMTC} & \textbf{eMBB} \\ \hline
Mobile Charging Robot               & x              &               &               \\ \hline
Charging for Charging Robot &                & x             &               \\ \hline
Teleoperated Driving                & x              &               & x             \\ \hline
Autonoumous Shuttle                 & x              &               & x             \\ \hline
Freight Robot                       & x              &               & x             \\ \hline
Smart Traffic Light                 & x              & x             &               \\ \hline
Camera                              & x               &               & x             \\ \hline
Electric Transportation                                &                & x             & x             \\ \hline
Bidirectional Charging     & x              &              & x              \\ \hline
Apps for User Interaction           &                &              & x              \\ \hline
\end{tabular}%
}
\end{table}

Table ~\ref{table:mapping} shows a classification between the proposed Smart city solutions and the major use cases of 5G based on the requirements of the pilots. Many pilots cannot be clearly mapped into one category. 

The mobile charging robot works largely autonomously. The driving functions and image processing for the identification of vehicles, as well as the evaluation of sensor information and the control of the actuators are carried out locally on the robot. It receives its charging orders from the energy control center via the 5G network and transmits the charging status and the charge transferred to the vehicle to the energy control center. The charging robot thus acts primarily like an IoT device but keeps always connectivity to its control center even with a low data rate. However, the robot moves around in freely accessible public areas where it inevitably comes into contact with people. Especially when it is traveling between its charging point and the public parking lot of the EVs, it passes heavily used paths. For safety reasons, an emergency shutdown is therefore required and constant remote control via the control center using uRLLC is necessary. 

The charging station of the charging robot mainly transmits measurement data on the charging process of the charging robot to the control center, therefore it can be considered as mMTC.  The terminal for bidirectional charging is a special case. In a V2G scenario, the battery of EVs are incorporated to the power grid as flexible and mobile energy storage to improve the usage of green energy and stabilize the power grid. As the energy flow can change direction briefly and repeatedly depending on the current energy production in the Ostra district and the load of the power grid, special requirements are placed on the latency and reliability of the control data. For this reason, the terminal for bidirectional charging is also considered a uRLLC case in addition to eMBB. 

Further uRLLC cases are presented by teleoperated driving, the autonomous shuttle and the freight robot. These pilots are controlled completely remotely by corresponding control centers. Camera streams and telemetry data must be transmitted with very low delays to ensure responsiveness. Safety-critical tasks are carried out using edge processing. Furthermore, these pilots use a camera system for object recognition and position determination. The transmission of the high-resolution streams requires high bandwidth and is therefore classified as an eMBB case.

The smart traffic lights of the Ostra district are primarily to be regarded as sensors, as they communicate status information on the traffic light phases and the remaining duration of the phase. As this data is relevant not only for traffic flow planning but also for teleoperated driving and autonomous shuttles, there are both mMTC and uRLLC requirements. 

In the Ostra district, camera systems are used to measure visitor flows and provide VRU detection for autonomous vehicles. Object recognition is outsourced to edge processing to limit the required bandwidth in the network. In addition, the results of the object detection are safety-critical information for autonomous vehicles. This is where requirements from eMBB and uRLLC meet. The electric transportation system (eBus) primarily collects static data on utilization. This is collected by evaluating sensor data and video streams.

Finally, the apps for user interaction will also be evaluated. The focus here is on the provision of status information. The user can order services such as charging an EV and receive corresponding information about charging. In addition, the utilization of sports facilities, ticket booking for events and transport options for arrival, departure and mobility within the Ostra district are to be provided. No special requirements are placed on the latency or reliability of the transmission, nor is a high bandwidth to be expected. Instead, the high number of people in the Ostra district alone can generate a large number of requests in the network. Nevertheless, it has to be classified as eMBB as it will be executed on regular smartphones that are fully 5G capable.  

\vspace{-1em}
\subsection{Enabler for Mobilities for EU}

5G has a unique feature set, to cover these diverse communication requirements of the pilots in a single communication network while ensuring that services do not disturb each other. In order to make use of the full potential of those features, the project is based on a stand-alone private 5G network with integrated edge processing capabilities as discussed in Section \ref{sec:KeyEnabler}. This section describes additional features of 5G that enable real-time communication and reliable connectivity between various project components and how the pilots benefit from them. 

\textbf{Network Slicing} allows each service to be granted a guaranteed bandwidth for its operation. For this purpose, the physical resources of the 5G network are divided into virtual network slices and assigned to a group of PDU sessions. At a logical level, the slices are completely separated from each other, which allows the network to be optimized individually for each application and also provides greater security against cyber attacks. Properties such as latency, bandwidth, reliability and security can  be configured, allowing 5G to isolate the pilots from each other and guarantee the respective requirements.

Pilots that exchange data streams with different characteristics can be coordinated using \textbf{Quality of Service (QoS)} in such a way that priorities are introduced. This allows data streams from a pilot to be handled individually, characteristics such as IP address, port, target data network or protocol type. 
In addition to the priority there is a packet delay budget and maximum packet error rate associated for each particular QoS stream to cover the specific requirements of the data stream.

Especially relevant for the uRLLC case, 5G has a resource type for QoS management that takes into account not only the packet error rate, a maximum bandwidth but also the shorter delay requirements of the packet. The so-called, \textbf{delay-critical PDU Sessions} introduces a maximum data burst volume in addition to the packet delay budget to describe the characteristics of cyclic traffic pattern which is typical for any kind of closed control loop. Using that concept, real-time data processing and transmission of packets can be enabled. 

\textbf{5G Reduced Capabilities (RedCap)} has been introduced as a light version of 5G with the goal of reducing complexity and limiting the bandwidth. This is important for small devices with limited communication requirements like wireless sensors in the mMTC use cases. To extend battery lifetime and improve the cost-efficiency of such devices optimization like narrower bandwidth, use of single transmit antenna and single receive antenna but also lower transmit power and limited modulation techniques take place in the system. 

On the other hand there is the eMBB service, which is focussing on high speed for user data and system capacity. This can be achieved by shifting the frequency spectrum towards higher areas such as cmWave and mmWave allowing higher bandwidth allocations. Furthermore, there are advanced antenna arrays enabling \textbf{massive MIMO}. A base station can support up to 256 antenna elements for intelligent beamforming and beam-tracking in the sub-6GHz band.    


\vspace{-1em}
\subsection{Integrated Architecture}
\begin{figure}[t]
 \centering
 \includegraphics[width=\linewidth]{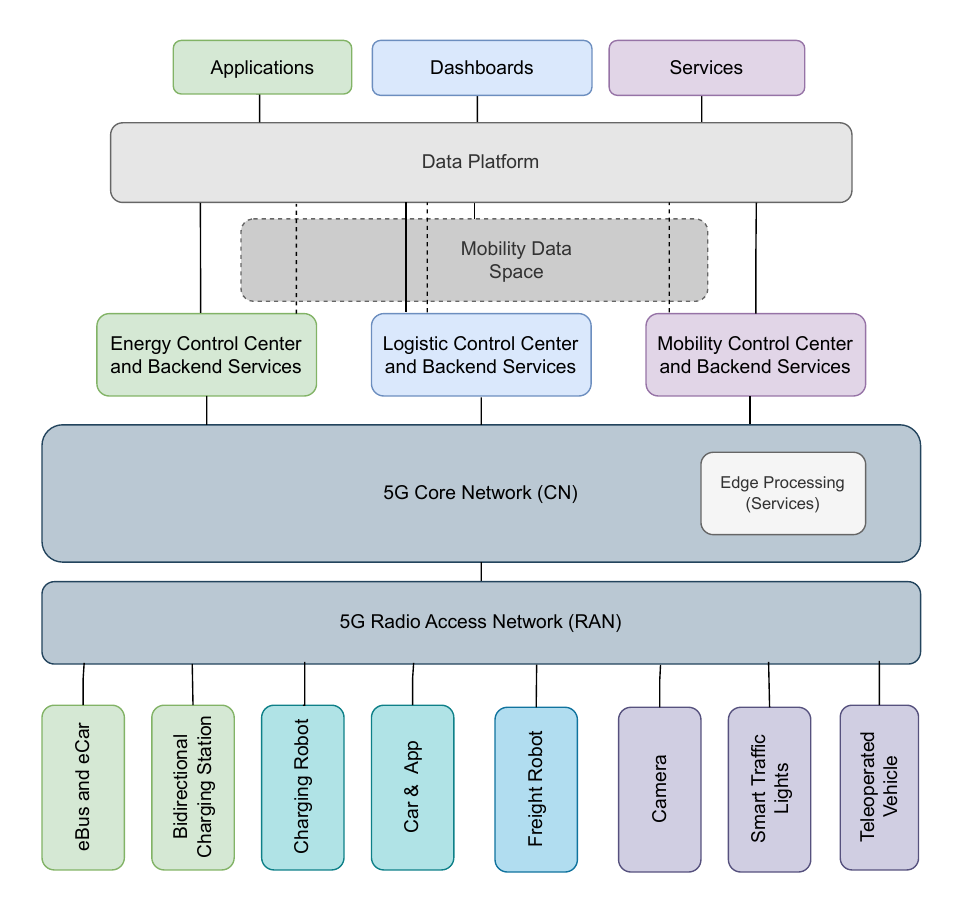}
 \caption{Integrated communication architecture for the private 5G network deployed in the Ostra district.}
 \label{fig:architecture}
\end{figure}

5G serves as a unifying technology for diverse smart city solutions, ensuring seamless connectivity and coordination, which can be seen in the architecture overview. 
At the heart of this integrated architecture lies a private 5G network, acting as the central communication backbone that interconnects various pilot projects within the Mobilities for EU initiative. Fig.~\ref{fig:architecture} illustrates this integrated architecture, highlighting the pivotal role of 5G in enabling seamless communication and coordination across different urban mobility solutions.

It connects active components such as mobile charging robots, and autonomous vehicles for transport of people and freight but also static devices such as cameras, traffic lights, bidirectional charging terminals with their respective counterpart control center, provides  edge processing capabilities for processing of control data and safety routines with uRLLC demands and finally ensures access to a comprehensive data platform. 

On a higher level the proposed pilots can be divided into three categories: energy, logistics and mobility optimization. Consequently, three different control centers are introduced to cover the fields. For sustainable aspects, there is a data platform connected to the control center, that provides statistics, trends and AI-based predictions for each control center. The mobility data space acts as digital rights management which ensures that the data exchange between the data platform, control center and third parties is based on predefined  agreements. In addition, it offers a dedicated interface, indicated with dashed lines, to securely exchange data between the data platform and control center using a specific tunnel within the mobility data space. 

The private 5G network connects the control center and the pilots while offering edge processing for image processing and control algorithms. With this integrated architecture in place, the next step is to evaluate the impact and sustainability of these 5G-enabled solutions.


\section{Sustainability Impact and Evaluation}

The Mobilities for EU project utilizes a comprehensive evaluation framework to assess the performance and efficiency of 2ZERO and CCAM solutions deployed in urban environments~\cite{mobilitiesforEU}. This section outlines our approach to assessing the impact of 5G technology, particularly through V2G technology, while addressing the unique challenges presented by small-scale pilots. V2G technology enables electric vehicles (EVs) to draw power from the grid but also return power to it, facilitating a bidirectional energy flow. This system is essential for balancing the grid, especially with the growing integration of renewable energy sources~\cite{rehman2023comprehensive}. By leveraging these advantages, the V2G pilot will be evaluated through a structured framework to assess its overall impact.

\vspace{-1em}
\subsection{V2G Implementation in Dresden's Ostra District}

Dresden's V2G pilot in the Ostra district aims to optimize energy usage, reduce peak load, and increase renewable energy integration. The system utilizes 5G's uRLLC and eMBB components, which offers several advantages:

\begin{itemize}
    \item Low Latency Communication: Enables real-time decision making for power flow between vehicles and the grid.
    \item High Reliability: Ensures consistent communication for critical grid balancing operations.
    \item Secure Connectivity: Protects sensitive data exchanges between EVs and the power grid.
    \end{itemize}

\vspace{-1em}
\subsection{Methodology}
The methodology for the Dresden V2G pilot project utilizes a phased approach to evaluate the impact of V2G technologies on energy consumption, cost savings, and CO2 emissions. 
Field tests are applied to validate the proposed technology, ensuring its effectiveness under real conditions. In addition, simulations are conducted to scale experiments to realistic real-world scenarios, providing valuable insights into how the technology performs in diverse load scenarios.
It features a co-simulation framework to model interactions between electric vehicles and the power grid, with real-time data collection for validation. Key data sources include performance metrics from the 5G network, which provides low latency, high reliability, and secure connectivity, ensuring effective operation of V2G systems within urban energy management frameworks.

\vspace{-1em}
\subsection{Framework for Evaluation}
To address the challenge of measuring 5G's impact, we employ the CIVITAS Initiative framework to assess the V2G pilot comprehensively across environmental, economic, transport, societal, and energy dimensions~\cite{engels2017CIVITAS}. This approach evaluates both technical performance and wider impacts on urban sustainability~\cite{mobilitiesforEU}.

\vspace{-1em}
\subsection{Categories of Impact Assessment for V2G}
We will focus on key categories that reflect the project's objectives and align with the European Green Deal. Table \ref{tab:my-table} outlines these impact assessment categories and metrics. This evaluation will enhance our understanding of how the 5G-enabled V2G system contributes to sustainable urban mobility, measuring both direct impacts and broader implications for energy efficiency and urban livability.

\begin{table}[]
\caption{Impact Assessment Categories for V2G Dresden Pilots}
\label{tab:my-table}
\begin{tabular}{|p{2cm}|p{6cm}|}
\hline
\multicolumn{1}{|c|}{\textbf{Category}} & \multicolumn{1}{c|}{\textbf{Impact Assessment Metrics}}                                                              \\ \hline
\textbf{Environment}  & \begin{tabular}{p{6cm}}Reduction in CO2 emissions\\ Reduction in NOx emissions\end{tabular}  \\ \hline
\textbf{Energy} &
  \begin{tabular}{p{6cm}}
  Energy consumption\\ Energy saving\\ Use of clean energy sources\\ 5G enhanced efficiency of V2G system through real-time data exchange between EVs and power grid\end{tabular} \\ \hline
\textbf{Transport}                      & \begin{tabular}{p{6cm}}Mileage\\ Charging times\end{tabular}                                                     \\ \hline
\textbf{Social}                         & \begin{tabular}{p{6cm}}Public acceptance of new technologies\\ Awareness of new solutions\end{tabular}           \\ \hline
\textbf{Economic}                       & \begin{tabular}{p{6cm}}Average operating costs\\ Pollution costs avoided\\ Overall economic impact\end{tabular} \\ \hline
\end{tabular}%
\end{table}

\vspace{-1em}
\subsection{Key Performance Indicators (KPIs)}
To measure the success of the interventions, we will develop a KPI matrix aligned with impact assessment categories. The KPIs will focus on quantifiable metrics reflecting the effectiveness of implemented solutions and their contribution to sustainability, including peak load reduction, renewable energy use, energy cost savings, grid stability, and user satisfaction. 

Linking these KPIs to the impact assessment categories will clarify how 5G technology supports the objectives of the Mobilities for EU project and aligns with the European Green Deal. These KPIs will also align with green 5G communication goals, emphasizing energy efficiency (EE), spectral efficiency (SE), and cost-effectiveness. The International Telecommunication Union (ITU) specifies KPIs for 5G networks~\cite{series2015imt, slalmi2021toward}, which include:
\begin{itemize}
    \item Fiber-like access data rates up to 20 Gbps
    \item Latency of mere milliseconds
    \item User Equipment Data Rates (UEDRs) up to 100 Mbps
    \item Connection density of up to 1 million devices/km²
\end{itemize}
To evaluate the impact of 5G on these KPIs, we will conduct before-and-after assessments and scenario analyses using the V2G pilot as a case study. By measuring latency, data rates, and connection density pre- and post-5G deployment, we can identify improvements. Simulations will also assess performance under varying load conditions, offering insights into how 5G enhances V2G systems and urban energy management across all pilots.

\section{Future Work and Challenges}
Despite the promising framework, challenges remain in quantifying the KPIs and establishing robust methodologies for measurement. It is essential to further investigate the specific role of 5G in the evaluation process to fully understand its contributions to sustainability outcomes.

Using V2G as a case study within the context of the Mobilities for EU project allows us to assess the sustainability impacts of various interventions while highlighting the enabling role of 5G technology. This approach not only informs future urban mobility initiatives but also contributes to achieving the ambitious climate goals set forth by the European Green Deal. The methodology developed for the V2G pilot will serve as a model for evaluating all other pilots within the project. Addressing these challenges will be vital for maximizing the potential of 5G technology in fostering sustainable urban environments.

Future research should focus on refining the measurement techniques for KPIs, exploring the long-term impacts of 5G-enabled solutions, and ensuring equitable access to these advancements. Addressing these challenges will be vital for maximizing the potential of 5G technology in fostering sustainable urban environments.

\section{Conclusion}
This paper has explored the transformative potential of the Mobilities for EU project, particularly through the lens of 5G technology and its application in the V2G pilot. By leveraging the CIVITAS framework, we systematically assessed the interventions aimed at enhancing urban mobility in Dresden, with a specific focus on sustainability metrics aligned with the European Green Deal.

The findings highlight that 5G technology serves as a critical enabler for smart urban mobility solutions, facilitating real-time data exchange and enhancing the efficiency of various mobility systems. The integration of innovative solutions, such as mobile charging robots, automated connected driving, and electrified public transport, demonstrates the effectiveness of 5G in addressing key urban challenges, including reducing CO2 and NOx emissions, improving energy efficiency, and promoting public acceptance of new technologies.

Moreover, the development of a comprehensive KPI matrix allows for the measurement of success across multiple dimensions, including environmental, economic, and social impacts. This structured approach not only provides insights into the effectiveness of the implemented solutions but also underscores the importance of 5G in achieving broader sustainability goals.

As cities continue to evolve towards smart urban environments, the lessons learned from the Mobilities for EU project can inform future initiatives. The potential of 5G technology to enhance urban mobility solutions is vast, yet challenges remain in quantifying its contributions and ensuring equitable access to these advancements.

In conclusion, the integration of 5G technology into urban mobility frameworks presents significant opportunities for creating sustainable, efficient, and livable cities. Continued research and collaboration among stakeholders will be essential to harness the full potential of these technologies, ultimately paving the way for a greener and more connected urban future.

\section*{Acknowledgement}
This work is funded by the European Union under Grant Agreement No 101139666, MOBILITIES FOR EU. Views and opinions expressed are those of the authors only and do not necessarily reflect those of the European Union or the European Climate, Infrastructure and Environment Executive Agency (CINEA). Neither the European Union nor the granting authority can be held responsible for them. It is supported by the German Research Foundation (DFG) as part of Germany's Excellence Strategy—EXC 2050/1—Cluster of Excellence “Centre for Tactile Internet with Human-in-the-Loop” (CeTI) of Technische Universität Dresden under project ID 390696704 and the Federal Ministry of Education and Research (BMBF) in the program of “Souverän. Digital. Vernetzt.” Joint project 6G-life, grant number 16KISK001K.

\bibliographystyle{IEEEtran}
\bibliography{IEEEabrv,references}
\end{document}